\documentclass[aps,prl,twocolumn,groupedaddress,showpacs]{revtex4}

\usepackage[bookmarks]{hyperref}
\usepackage[dvips]{graphicx}
\usepackage{amsmath}
\usepackage{amstext}
\usepackage{graphics}
\usepackage{exscale}
\usepackage{epsfig}
\usepackage[T1]{fontenc}
\usepackage{bm}
\usepackage{bbm}
\usepackage{color}

\usepackage{version}

\usepackage{latexsym}

\usepackage[tight]{subfigure}

\renewcommand{\epsilon}{\varepsilon}

\newcommand{\integral}[3]{\!\int\limits_{#2}^{#3}\!\!{\rm d}#1\;}

\newcommand{\elcre}[2]{ c^{\dagger}_{#1,#2}}
\newcommand{\elann}[2]{ c_{#1,#2}}

\newcommand{\vq}{{\bm q}}

\newcommand{\hc}{\mathrm{h.c.}}

\includeversion{commentst1}

\begin{document}

\title{Theory of electron-phonon superconductivity: Does retardation really lead to a small Coulomb pseudopotential?}
\author{Johannes Bauer${}^1$, Jong E. Han${}^{1,2}$, and Olle
Gunnarsson${}^1$}
\affiliation{${}^1$Max-Planck Institute for Solid State Research,
Heisenbergstr.1,
 70569 Stuttgart, Germany}
\affiliation{${}^2$  Department of Physics, SUNY at Buffalo, Buffalo,
New York 14260, USA}
\date{\today}
\begin{abstract}
The theory of electron-phonon superconductivity depends on retardation drastically 
reducing effects of the strong Coulomb repulsion. The standard theory 
only treats the lowest order diagram, which is an uncontrolled approximation. 
We study retardation in the Hubbard-Holstein model in a controlled way using perturbation 
theory and dynamical mean-field theory. We calculate analytically second order
results for the pseudopotential $\mu^*$ and demonstrate the validity up to
intermediate couplings by comparison with non-perturbative
results. Retardation effects are still operative, but less 
efficient, leading to somewhat larger values of $\mu^*$. Therefore, our theory can 
help to understand situations where the standard theory yields
overestimates for $T_c$.
\end{abstract}

\maketitle


The theory of superconductivity based on the electron-phonon mechanism
has been very successful in describing the properties of many materials
\cite{Car90,MC08}.
The electron-phonon coupling is treated in the Migdal Eliashberg
(ME) theory \cite{Eli60}, which relies on Migdal's theorem \cite{Mig58}.
This employs the fact that typical electron ($E_{\rm el}$) and phonon
($\omega_{\rm  ph}$) energy scales differ largely. Then
perturbation theory greatly simplifies as vertex corrections are
small. This is true even for large values of the electron-phonon coupling
parameter $\lambda>1$ as long as $\lambda \omega_{\rm   ph}/E_{\rm el}$
remains small \cite{Mig58,MK82,BHG11}.

A crucial issue is the effect of the Coulomb repulsion, 
typically much larger than the phonon-induced attraction.
The electronic repulsion in the pairing channel can be projected to
the phonon scale. It is then strongly reduced due to retardation effects and
one finds     \cite{BTS62,MA62,schrieffer,Sca69},
\begin{equation}
  \mu_c^*=\frac{\mu_c}{1+\mu_c\log\Big(\frac{E_{\rm el}}{\omega_{\rm
        ph}}\Big)},
\label{eq:mustar}
\end{equation}
often termed the Morel-Anderson (MA) pseudopotential.
Here, $\mu_c=\rho_0 U$, where $U$ is a typical screened Coulomb interaction
and $\rho_0$ is the density of states (DOS) at the Fermi energy.
Since usually $E_{\rm el} \gg \omega_{\rm ph}$, one finds that $\mu_c^* \ll \mu_c$
and often also $\mu_c^* <  \lambda$.
Eq. (\ref{eq:mustar}) leads to estimates of the order
$\mu_c^*\sim 0.1-0.14$. This agrees rather well to the fitting parameter $\mu^*$
obtained from tunnelling spectroscopy for many conventional superconductors
 \cite{Car90}.

Although ME theory has been very successful, the
treatment of the Coulomb repulsion is by no means rigorous \cite{AM82}.
For the electron-phonon interaction, Migdal's theorem justifies the neglect of
vertex corrections. For the Coulomb interaction there is no similar
justification, and Eq.~(\ref{eq:mustar}) is based  
on an uncontrolled approximation. As long as it has not been demonstrated 
that $\mu_c^*$ indeed is small, conventional superconductivity has not 
been properly explained. The purpose of this paper is to analyse higher order
corrections to the MA result.  We show that retardation
effects also reduce higher order contributions beyond Eq.~(\ref{eq:mustar}), 
although less efficiently. For moderate $\mu_c$ we then find that $\mu_c^*$ 
indeed is rather small, although somewhat larger than in the standard theory. 

There are cases reported in the literature, e.g., V or Nb${}_3$Ge 
\cite{Car90}, where the experimental values for $\mu^*$ in the literature of
the order $0.2-0.3$ substantially larger 
than the traditional quotes, even though the ratio $E_{\rm el}/\omega_{\rm
  ph}$ is not much different. These are not well explained by
Eq.~(\ref{eq:mustar}). Density functional theory (DFT) \cite{SS96} finds good
agreement with the tunnelling results for the pairing function, but to explain
the experimental values for $T_c$, in some cases quite large values of
$\mu^*$ have to be used . A prominent example is elemental Li at ambient
pressure \cite{LC91,RA97}, where the coupling constant was estimated to be
$\lambda\sim 0.4$ \cite{LC91,BNC11}. With $\mu^*\sim 0.1$  
this implies $T_c\sim$~1K, while experimentally
$T_c\sim$~0.4mK\cite{TJUPS07}, which requires $\mu^*\sim 0.23$. The role of
the Coulomb pseudo potential was also discussed in the case of the
alkali-doped fullerides. Here, the MA theory leads to a large reduction of 
$\mu_c^{\ast}$ due to couplings to higher sub-bands \cite{GZ92}, although this
is unphysical \cite{And91pre,GZ92} 
and raises serious questions about higher order corrections for           
molecular solids. Actually, it was found that superconductivity
in fullerides is due to a complicated interplay between the Coulomb
interaction and Jahn-Teller phonons \cite{CFCT02,HGC03}.

The MA theory corresponds to treating the irreducible vertex
to first order in $U$. Berk and Schrieffer \cite{BS66} included a specific
class of higher order diagrams describing the coupling to ferromagnetic (FM) spin
fluctuations, addressing almost FM metals, like Pd. They   
found that retardation is ineffective for the added diagrams 
and that superconductivity is strongly suppressed, which can help to explain
the cases when FM spin fluctuations are important. 
For the model considered below, we extend the MA approach 
by adding the second order term. 
This does not include a large enhancement of the spin susceptibility and
therefore we address the large class of systems which are not close to a FM instability.
We use a projection approach 
and provide numerical calculations of $\mu_c^{\ast}$ without further
approximation, as well as approximate analytical calculations.
We find
that retardation effects lead to a reduction of $\mu_c\to \mu_c^{\ast}$
also in the second order calculation, but less efficiently. 
Then we add the electron-phonon interaction and
calculate the superconducting gap numerically and approximately
analytically. The results are well understood by the derived results
for $\mu_c^{\ast}$.  To check the range of validity, our
calculations are compared with
non-perturbative dynamical mean-field theory (DMFT), which include all higher
order corrections. DMFT fully treats retardation effects, which are crucial
for the Coulomb pseudopotential. We show that results based on the 
perturbation theory agree well with DMFT calculations up to
intermediate couplings $\mu_c\sim 0.5$.

In this work we deal with generic features of electron-phonon
superconductivity and do not carry out calculations for a specific material.
We employ the Hubbard-Holstein model, which possesses all necessary
ingredients,
\begin{eqnarray}
  \label{hubholham}
  H&=&-\sum_{i,j,{\sigma}}(t_{ij}\elcre i{\sigma}\elann
j{\sigma}+\hc)+U\sum_i\hat n_{i,\uparrow}\hat n_{i,\downarrow} \\
&&+\omega_0\sum_ib_i^{\dagger}b_i+g\sum_i(b_i+b_i^{\dagger})\Big(\sum_{\sigma}\hat
n_{i,\sigma}-1\Big).
\nonumber
\end{eqnarray}
$\elcre i{\sigma}$ creates an electron at site $i$ with spin $\sigma$,
and $b_i^{\dagger}$ a phonon with oscillator frequency $\omega_0$,
$\hat n_{i,\sigma}=\elcre i{\sigma}\elann
i{\sigma}$. The electrons interact locally with a screened Coulomb interaction $U$, and
couple to an optical phonon  with coupling constant
$g$.  For infinite  dimensions this model is  solved
exactly by the DMFT.

First we deduce $\mu_c^*$ from an analysis of the pairing instability,
in the limit $i\omega_n\to 0$, $\vq\to 0$. We define the symmetric
matrix \cite{GKKR96},
\begin{equation}
A_{n,m}=\delta_{n,m}- \frac{1}{\beta}\sqrt{\tilde\chi^0(i\omega_{n})}\Gamma^{(\rm
  pp)}(i\omega_{n},i\omega_{m};0)\sqrt{\tilde\chi^0(i\omega_{m})},
\label{symmatrixinst}
\end{equation}
where $\beta$ is the inverse temperature, $\Gamma^{(\rm  pp)}(i\omega_{n_1},i\omega_{n_2};i\omega_n)$ is the irreducible
vertex in the particle-particle channel and the pair propagator is   
$\tilde\chi^0(i\omega_{n})=[G(i\omega_{n})-G(-i\omega_{n})]/[\zeta(-i\omega_{n})-\zeta(i\omega_{n})]$,
where $\zeta(i\omega_{n})=i\omega_{n}+\mu-\Sigma(i\omega_{n})$ and
$G(i\omega_{n_1})$ is the local lattice Green's function.
$A$ is      singular at $T_c$.  We introduce the ``low-energy part''
\begin{equation}
\label{eq:o0}
  A^{\rm low}_{nm}=A_{nm}-\sum_{|\omega_{n'}|,|\omega_{m'}|>\omega_{\rm ph}}A_{nn'}[\bar A^{-1}]_{n'm'}A_{m'm},
\end{equation}
$n,m$ such that $|\omega_n|,|\omega_m|<\omega_{\rm ph}$, and $\bar A$ is the
the block for $|\omega_n|,|\omega_m|>\omega_{\rm ph}$. If $A^{\rm low}$ is singular, $A$ is also singular.
The ``folding in'' of larger frequencies describes how retardation effects
reduce effects of the Coulomb repulsion on low frequency properties.
We first consider the lowest order term of $\Gamma^{(\rm pp)}$ in $U$,
$\Gamma^{(\rm pp),1}=-U$.
We focus on the dependence on the half-band width $D$ and assume
a constant DOS, $\rho_0=1/(2D)$.
It is  a rather good approximation to write
$\tilde \chi^0(i\omega_n)=
\rho_0\pi/|\omega_n|$,
if $|\omega_n|<D$ and 0 otherwise.
With $\mu_c=\rho_0 U$, $A$ takes the form,
\begin{equation}
\label{eq:o1}
A_{nm}=\delta_{nm}+\frac{\pi }{\beta \sqrt{|\omega_n \omega_m|}}\mu_c,
\end{equation}
which is separable and can be inverted exactly.
Replacing summations by integrals, we find then
\begin{eqnarray}
\label{eq:o3}
&&A_{nm}^{\rm low}=\delta_{nm}+\frac{\pi }{\beta \sqrt{|\omega_n \omega_m|}}
\frac{ \mu_c}{1+\mu_c{\rm log}(D/\omega_{\rm ph})}.
\end{eqnarray}
Comparison of Eq.~(\ref{eq:o1}) and (\ref{eq:o3}) leads to the Coulomb
pseudopotential, $\mu_c\to\mu_c^*$, as given in in
Eq.~(\ref{eq:mustar}).

We next consider the second order term of $\Gamma^{(\rm pp)}$ in $U$,
which comes from a crossed diagram,
\begin{equation}
\label{eq:o00}
  \Gamma^{(\rm pp),2}(i\omega_{n_1},i\omega_{n_2};0)= U^2 \Pi(i\omega_{n_1}+i\omega_{n_2}),
\end{equation}
where the particle-hole bubble is given by
\begin{equation}
\Pi(i\omega_n)= \frac{1}{\beta} \sum_{m} G(i\omega_n+i\omega_{m})G(i\omega_{m}).
\label{pidef}
\end{equation}
We can write $\Pi(i\omega_n)=-f(x)a\rho_0$, $x=i\omega_n \rho_0$, where $f$ is
independent of $D$ and approximated as
\begin{equation}
\label{eq:o6}
f(x)=\frac{1}{1+b|x|+cx^2},
\end{equation}
where $a=1.38$, $b=2$ and $c=5$ are suitable values for the constant DOS.

Because of the form of $\Pi(i\omega_n+i\omega_m)$, $\bar A$ in Eq.~(\ref{eq:o0})
cannot be inverted analytically. Instead we use the inverse of $\bar A$
based on $\Gamma^{(\rm pp),1}$, now only correct to first order in $U$.
However, since the off-diagonal terms of $A$ in Eq.~(\ref{eq:o0}) are of
order $U$, the final analytical result is correct to order $U^3$.
We make an ansatz for $\mu_c^*$ similar to Eq. (\ref{eq:mustar}),
\begin{equation}
  \mu_c^*=\frac{\mu_c+a\mu_c^2 }{1+\mu_c \log\Big(\frac{D}{\omega_{\rm
        ph}}\Big)+a\mu_c^2 \log\Big(\frac{\alpha D}{\omega_{\rm ph}}\Big)}.
\label{eq:mustarproj}
\end{equation}
Eq.~(\ref{eq:o6}) shows that the ``folding in'' of $\Gamma^{(\rm pp),2}$
for large frequencies  gives a small contribution
 to $A^{\rm low}$. This implies a reduced  effective band width for the second
order term, described by the                     factor $\alpha$ in the
logarithm. Identifying with the analytical result correct to order
$U^3$, $\alpha \approx 0.10$ is obtained.
Eq.~(\ref{eq:mustarproj}) is then also correct to order $U^3$.

\begin{figure}[!t]
\centering
\includegraphics[width=0.3\textwidth,angle=270]{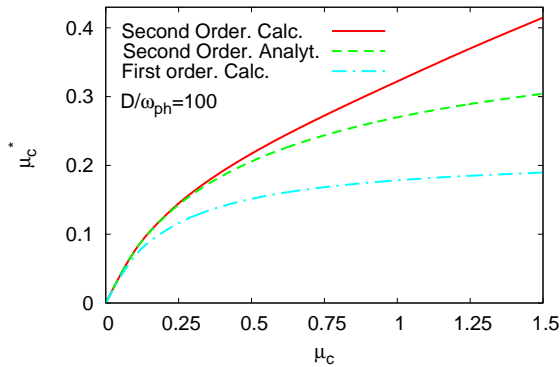}
\vspace*{-0.3cm}
\caption{(Color online) $\mu^*_c$ as a function of $\mu_c$ for $D/\omega_{\rm ph}=100$ and $\beta \omega_{\rm ph}=240$.
The figure shows the calculated results using both the first order and first plus second
order result for $\Gamma^{(\rm pp)}$ as well as the approximation in
Eq.~(\ref{eq:mustarproj}).  }
\label{fig:mustarproj}
\end{figure}
\noindent
Fig. \ref{fig:mustarproj} shows results obtained by performing the calculations
in Eq.~(\ref{eq:o0}) numerically using $\Gamma^{(\rm pp)}$ up to first or
second order in $U$  and  $\tilde\chi^0$  with $\Sigma(i\omega_n)\equiv 0$.
The analytical result in Eq.~(\ref{eq:mustarproj}) are also shown.
The second order result is clearly larger than the first order result. 
For $\mu_c \leq 0.5$,
Eq.~(\ref{eq:mustarproj}) describes the second order calculation rather well,
while for larger $\mu_c$ corrections to the analytic result
make $\mu_c^*$ still larger compared to Eq.~(\ref{eq:mustarproj}).
The second order contribution is reduced by retardation effects,
but it is substantially less efficient than for the first order contribution, 
as described by the  factor $\alpha\sim 0.1$.

The MA theory [Eq.~(\ref{eq:mustar})] makes two main
predictions: (i) as $D$ is increased for fixed $\mu_c$ and $\omega_{\rm
  ph}$,  $\mu^{\ast}_c$ goes to zero and (ii) as $\mu_c$ is increased for fixed $D/\omega_{\rm ph}$,
$\mu^{\ast}_c$ saturates at the value $1/{\rm log}(D/\omega_{\rm ph})$. (i) remains true
when the second order contribution $\Gamma^{(\rm pp),2}$ is taken into
account, but the numerical result in Fig.~\ref{fig:mustarproj} shows that (ii) is violated,
i.e, $\mu^{\ast}_c$ does not saturate as $U$ is increased.
Berk and Schrieffer \cite{BS66} found that retardation is ineffective for
higher order terms. They focused on almost
FM metals for which the spin susceptibility is strongly
enhanced for small ${\vq}$ and $\omega$. This corresponds to
low-lying excitations for which one expects small retardation
effects. This is different from the situation considered here.


As a complimentary analysis, we extract results
for $\mu_c^*$ from the spectral gap $\Delta_{\rm sp}$  
at $T=0$. This is similar to the original work by Morel and
Anderson \cite{MA62}, which included only the first order term
in $U$. We work on the imaginary axis in the limit $T\to 0$. Starting
point is the self-consistency equation for the off-diagonal
self-energy,
\begin{equation}
    \Sigma_{21}(i\omega_n)=\frac{1}{\beta}\sum_{m}G_{21}(i\omega_m)
    K(i\omega_n,i\omega_m) ,
\label{eq:selfconSig21}
\end{equation}
where the kernel $K(i\omega_n,i\omega_m)$ includes the attraction
mediated by the phonons and the repulsion to order $U^2$,
\begin{equation}
  K(i\omega_n,i\omega_m)=
-\frac{\lambda}{\rho_0}\frac{1}{1+\big(\frac{\omega_n-\omega_m}{\omega_{\rm
        ph}}\big)^2} 
+U  -U^2\Pi(i\omega_n+i\omega_m).
\label{eq:kernel}
\end{equation}
$G_{21}(i\omega_m)$ is the offdiagonal Green's function \cite{BHD09,BHG11}
and a semi-elliptic DOS $\rho_0(\epsilon)=\sqrt{4t^2-\epsilon^2}/(2\pi
t^2)$ with $W=4t=2D$ is used.
The effect of the diagonal self-energy $\Sigma_{11}$ is taken into account in the
analytical calculations by a factor $Z=1-\Sigma_{11}'(0)$, which is
taken finite only for $|\omega_n|<\omega_{\rm ph}$ [see Eq. (\ref{eq:g21})].

The self-consistency equation (\ref{eq:selfconSig21}) can be solved numerically
by iteration. For an analytical solution, we need to make some approximations.
At half filling,  we use for the Green's function for $|\omega_n|<\omega_{\rm ph}$,
\begin{equation}
  G_{21}(i\omega_n)\simeq -\frac{1}{t}
\frac{\Sigma_{21}(i\omega_n)}{\sqrt{Z^2\omega_n^2+\Sigma_{21}(i\omega_n)^2}}
\label{eq:g21}
\end{equation}
for $\omega_{\rm ph}<|\omega_n|<D$,
$G_{21}(i\omega_n)\simeq -\Sigma_{21}(i\omega_n)/(t|\omega_n|)$,
and for $|\omega_n|>D$, $G_{21}(i\omega_n)\simeq 0$.
A suitable ansatz for the off-diagonal self-energy is \cite{MA62},
\begin{equation}
  \Sigma_{21}(i\omega_n)=\Delta_3+ \Delta_2 f(i\omega_n\rho_0)+
\frac{\Delta_1-\Delta_2-\Delta_3}{1+\big(\frac{\omega_n}{\omega_{\rm ph}}\big)^2}.
\end{equation}
For $f(x)$ we use Eq. (\ref{eq:o6}), except that the numerical coefficients
are modified for the semi-elliptic DOS.  We have to solve for
the three parameters $\Delta_1$, $\Delta_2$, and $\Delta_3$ by evaluating the
self-consistency equation at suitable values of $i\omega$. The
general case is algebraically quite involved. Here, we only treat the
first and purely second order cases explicitly to see the major effects.

For the first order case, we set $\Delta_2=0$ and omit the
$U^2$-term in Eq.~(\ref{eq:kernel}). We use the conditions
$\Sigma_{21}(0)=\Delta_1$, and $\Sigma_{21}(iD) \simeq \Delta_3$.
and assume $\Delta_i\ll\omega_{\rm ph}\ll D$.
With the usual approximations we find a solution for the spectral gap
$\Delta_{\rm sp}=\Delta_1/Z$ of the form \cite{McM68,AD75},
\begin{equation}
  \Delta_{\rm sp} = c_1 \omega_{\rm ph}\exp\Big(-\frac{Z
      c_2}{\lambda-\mu_c^*(1+c_3\lambda)-\mu_{c,1}^*}\Big).
\label{eq:gapanafitform}
\end{equation}
The result for $\mu_c^*$ is given in Eq. (\ref{eq:mustar}) and $\mu_{c,1}^*=0$.

In the situation when only the $U^2$-term is included we set
$\Delta_3=0$ and omit the constant $U$-term in Eq.~(\ref{eq:kernel}).
To determine the parameters $\Delta_1$ and $\Delta_2$, we use the
following two conditions: $\Sigma_{21}(0)=\Delta_1$,
and $\Sigma_{21}(iD)+\Sigma_{21}(-iD) \simeq 2\Delta_2 f(i/2)$.
The calculation again yields a result of the form
(\ref{eq:gapanafitform}), however, now with
\begin{equation}
  \mu_c^*=\frac{a\mu_c^2 }{1+a\mu_c^2 \log\Big(\frac{\alpha_2 D}{\omega_{\rm
        ph}}\Big)},\; \mu_{c,1}^*=\frac{\gamma a^2\mu_c^4}{1+a\mu_c^2 \log\Big(\frac{\alpha_2 D}{\omega_{\rm
        ph}}\Big)}.
\label{eq:mustarho}
\end{equation}
As before $0<\alpha_2 < 1$ accounts for the less effective retardation
effects.  In addition a term $\mu_{c,1}^*$ appears,
which was absent in the first order calculation. Such terms can account
for the discrepancy between analytical and numerical results
in Fig. \ref{fig:mustarproj}, where the analytical result 
saturates as function of $\mu_c$. We obtain           
 $\gamma\approx 0.8$ for $D\gg\omega_{\rm ph}$.

\begin{figure}[!t]
\centering
\includegraphics[width=0.42\textwidth]{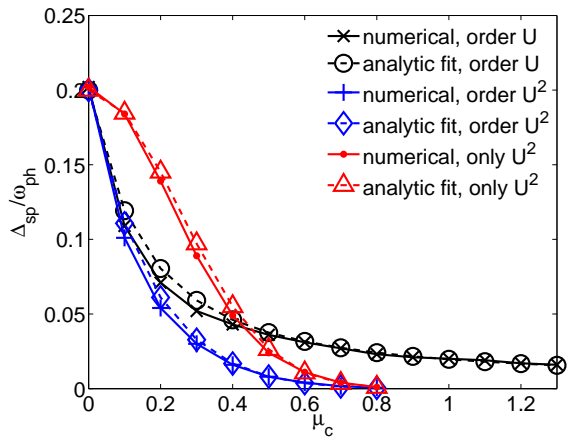}
\vspace*{-0.3cm}
\caption{(Color online) The spectral gap $\Delta_{\rm sp}$ as calculated from the numerical solution
  of Eq. (\ref{eq:selfconSig21}) with different kernels as a  function
  of $\mu_c$ for $\lambda=0.5$ and $D/\omega_{\rm  ph}=80$ in comparison with the
  corresponding analytical results based on Eq. (\ref{eq:gapanafitform}).}
\label{fig:gapmucdepUU2onlyU}
\end{figure}
\noindent
In Fig. \ref{fig:gapmucdepUU2onlyU}, we show the $\mu_c$-dependence of
the numerical solution of Eq. (\ref{eq:selfconSig21}) for $\lambda=0.5$. It is compared
with the analytical result in Eq. (\ref{eq:gapanafitform}) with the
respective results for $\mu_c^*$,
Eqs. (\ref{eq:mustar},\ref{eq:mustarproj},\ref{eq:mustarho}). We use
$\alpha$ as before and $\alpha_2\approx 0.2$.
$c_1=1.7$, $c_2=1.07$ were determined by fitting to the numerical solution for
$\mu_c=0$ in the regime $0<\lambda<0.5$. $c_3=0.8$ was found to give
a reasonable fit for the first order calculation in $\mu_c$. We omit
the term $\mu_{c,1}^*$ for the values of $\mu_c$ considered.
The agreement between numerical and analytical results is quite good,
which supports the earlier findings from the projection approach
Eq. (\ref{eq:o0}). One finds similar results as in
Fig. \ref{fig:gapmucdepUU2onlyU}  when calculating $T_c$.


\begin{figure}[!t]
\centering
\includegraphics[width=0.42\textwidth]{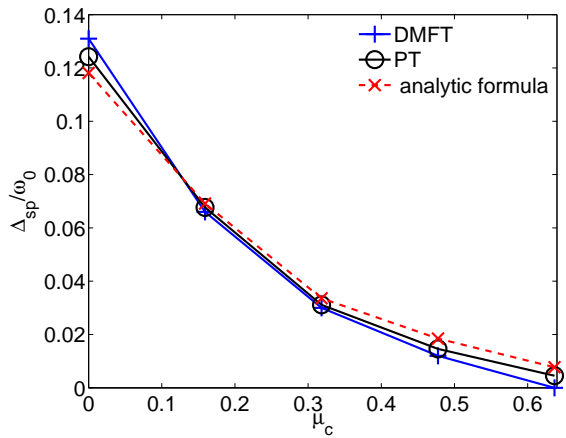}
\vspace*{-0.3cm}
\caption{(Color online) DMFT result for the spectral
  gap $\Delta_{\rm sp} \simeq z\Sigma^{\rm off}(0)$ as a function of $\mu_c$ for $D=2$, constant
  $\lambda\simeq 1$ according to the second order result for
  $g^r$ in Eq. (\ref{eq:effvertcorr2}) and Eq. (\ref{lambdafinU}), and
  $\omega_0^r\simeq 0.05$ in comparison with the PT. We also included the result of the analytic formula in
  Eq. (\ref{eq:gapanafitform}) with renormalized parameters
  $\omega_{\rm ph}=\omega_0^r$, $Z$, calculated from PT, $\mu_c^*$ from
  Eq. (\ref{eq:mustarproj}) and $c_i$ as in Fig. \ref{fig:gapmucdepUU2onlyU}.}
\label{fig:gapconslambdar_mucdep}
\end{figure}
\noindent
We now want to corroborate our findings with DMFT and analyse the
impact of increasing $\mu_c$ on
superconductivity similar to Fig. \ref{fig:gapmucdepUU2onlyU}.
For this purpose we need to include additional effects. We extended our
previous DMFT and ME perturbation theory (PT) \cite{BHG11} to the case
of finite $U$. In the PT for $\Sigma_{11}$ and $\Sigma_{21}$ we include
direct terms in $U$ up to second order, for instance, the terms
described in Eq. (\ref{eq:selfconSig21}). In addition, we have to take
into account the fact that the electron-phonon vertex $\Gamma^{(\rm
  ep)}$ is renormalized by the Coulomb repulsion. We
introduce the quantity $\Gamma^{(\rm  ep)}_U$, which contains $g$ and
all corrections from the $U$-term. For a weak frequency dependence up
to the small phonon scale, we can define a renormalized coupling $g^r
= \Gamma^{(\rm  ep)}_U(0,0)$.
For $g^r$ we use the result up to second order,
\begin{equation}
  \frac{g^r}{g}=1-a_1\mu_c+(a_1^2-a_2)\mu_c^2,
\label{eq:effvertcorr2}
\end{equation}
where $a_1=-\Pi(0)/\rho_0$ and $a_2=\frac{1}{\rho_0^2\beta}
\sum_{k}\Pi(i\omega_k)G(i\omega_k)^2$.
This was found to give a good description up to $U\sim D$
\cite{HHAS03}. Thus for the electron phonon part we use
\begin{equation}
\underline{\Sigma}^{\rm el-ph}(i\omega_n)=
-\frac{1}{\beta}\sum_{m}[g^r]^2 \tau_3
\underline{G}(i\omega_m+i\omega_n)\tau_3 D(i\omega_m) ,
\label{MEgreq}
\end{equation}
where the phonon propagator $D(i\omega_m)$ is taken from the DMFT
calculation \cite{BHG11}. The effective $\lambda$ is defined by
\begin{equation}
  \lambda=2\rho_0[g^r]^2\integral{\omega}{0}{\infty}\frac{\rho^D(\omega)}{\omega},
\label{lambdafinU}
\end{equation}
where the phonon spectral function $\rho^D(\omega)$ includes
self-energy corrections due to $U$. We define the
renormalized phonon energy scale $\omega_{\rm ph}=\omega_0^r$ by the peak
position of $\rho^D(\omega)$.
 Through the condition $\lambda\simeq 1$ and $\omega_{\rm
  ph}=\omega_0^r\simeq 0.05$ a set of bare model parameters
$(g,\omega_0,\mu_c)$ is determined by the DMFT calculations,
for which we can compare the PT with DMFT.
The results are shown in Fig.~\ref{fig:gapconslambdar_mucdep}.

We find good agreement of the DMFT result with PT and the
analytic formula, Eq. (\ref{eq:gapanafitform}), up to $\mu_c\sim 0.4-0.5$. This demonstrates that (i) the
electron-phonon vertex correction according to
Eq. (\ref{eq:effvertcorr2}) is suitable, (ii) that the
higher order form for the Coulomb pseudopotential in
Eq. (\ref{eq:mustarproj}) captures correctly the
results of the PT and the full DMFT calculation, and (iii) that the
effective parameter description is appropriate.  Therefore, this
validates the previous analysis in a more complete calculation and it
corroborates our findings for $\mu_c^*$ up to intermediate values of
$\mu_c$.
For larger values of $\mu_c$, we find that $\Delta_{\rm sp}$ in the
PT calculations exceeds the DMFT result, where $\Delta_{\rm sp}\to 0$.
Then both Eq. (\ref{eq:effvertcorr2}) for $g^r$ and
Eq. (\ref{eq:mustarproj}) for $\mu_c^*$  start to underestimate the
reduction effect.


In conclusion, we emphasise that the standard theory of how retardation
reduces $\mu_c\to \mu_c^{\ast}$ is based on an uncontrolled approximation,
since there is no Migdal's theorem for the Coulomb interaction. In a
controlled framework we analyse higher order corrections. We obtain an
analytical expression for the next order term, and show that retardation also
reduces this term, however substantially less efficiently. Non-perturbative
DMFT calculations demonstrate that the perturbative result is accurate up 
to intermediate couplings. The main conclusion is then that retardation effects
indeed lead to rather small values of $\mu_c^*$, even when contributions
beyond the standard theory are considered. For systems with sizable Coulomb
interactions $\mu_c$, our values for $\mu_c^*$ are larger than in the
standard theory and lead to reduced values of the superconducting gap and $T_c$. 
We have focused on the reduction of phonon induced s-wave
superconductivity due to the Coulomb repulsion between
electrons. Superconductivity which is induced in an anisotropic higher order
angular momentum channel by purely repulsive interactions, such as the well-known
Kohn-Luttinger effect \cite{KL65}, is not dealt with in the present work.    

\paragraph{Acknowledgment -}
We wish to thank N. Dupuis, A.C. Hewson, P. Horsch, C. Husemann,
D. Manske, and  R. Zeyher for helpful discussions. JH acknowledges
support from the grant NSF DMR-0907150.

\bibliography{artikel,biblio1}

\end{document}